\documentclass[epj]{webofc}
\usepackage[utf8]{inputenc}
\usepackage[varg]{txfonts}   
\usepackage{booktabs}
\usepackage{xcolor}
\definecolor{darkred}{rgb}{0.4,0.0,0.0}
\definecolor{darkgreen}{rgb}{0.0,0.4,0.0}
\definecolor{darkblue}{rgb}{0.0,0.0,0.4}

\newcommand\tildeG{{\widetilde G}}

\usepackage[bookmarks,linktocpage,colorlinks,
    linkcolor = darkred,
    urlcolor  = darkblue,
    citecolor = darkgreen]{hyperref}

\usepackage{subfigure}
\wocname{EPJ Web of Conferences}
\woctitle{Lattice2017}
\newcommand{\ecm}{e{\cdot}\textrm{cm}}
\begin{document}
%
\selectlanguage{english}
\title{%
Neutron Electric Dipole Moment on the Lattice
}
\author{%
\firstname{Boram} \lastname{Yoon}\inst{1}\fnsep\thanks{Speaker, \email{boram@lanl.gov} } \and
\firstname{Tanmoy} \lastname{Bhattacharya}\inst{1} \and
\firstname{Rajan}  \lastname{Gupta}\inst{1}
}
\institute{%
Los Alamos National Laboratory, Los Alamos, NM 87545, USA
}
\abstract{%
  For the neutron to have an electric dipole moment (EDM), the theory
  of nature must have T, or equivalently CP, violation. Neutron EDM 
  is a very good probe of novel CP violation in
  beyond the standard model physics. To leverage the connection
  between measured neutron EDM and novel mechanism of CP violation, one
  requires the calculation of matrix elements for CP violating operators, 
  for which lattice QCD provides a first principle method. In this paper,
  we review the status of recent lattice QCD calculations of the
  contributions of the QCD $\Theta$-term, the quark EDM term, and the
  quark chromo-EDM term to the neutron EDM.  }
\maketitle

\section{Introduction}\label{intro}

Electric dipole moment (EDM) of the neutron measures the separation of
positive and negative charge in the neutron and is necessarily aligned
along the spin axis.  For a neutron to have an EDM, the theory of
elementary particles must violate parity (P) and time reversal (T)
invariance, and charge conjugation and parity (CP) symmetry if CPT is
conserved. Under a parity transformation, the EDM of any particle
changes its sign while the direction of the spin remains the
same. Under time-reversal transformation, the spin flips its direction
while the EDM stays unchanged.

\begin{figure}[thb]
  \centering
  \includegraphics[width=0.7\textwidth]{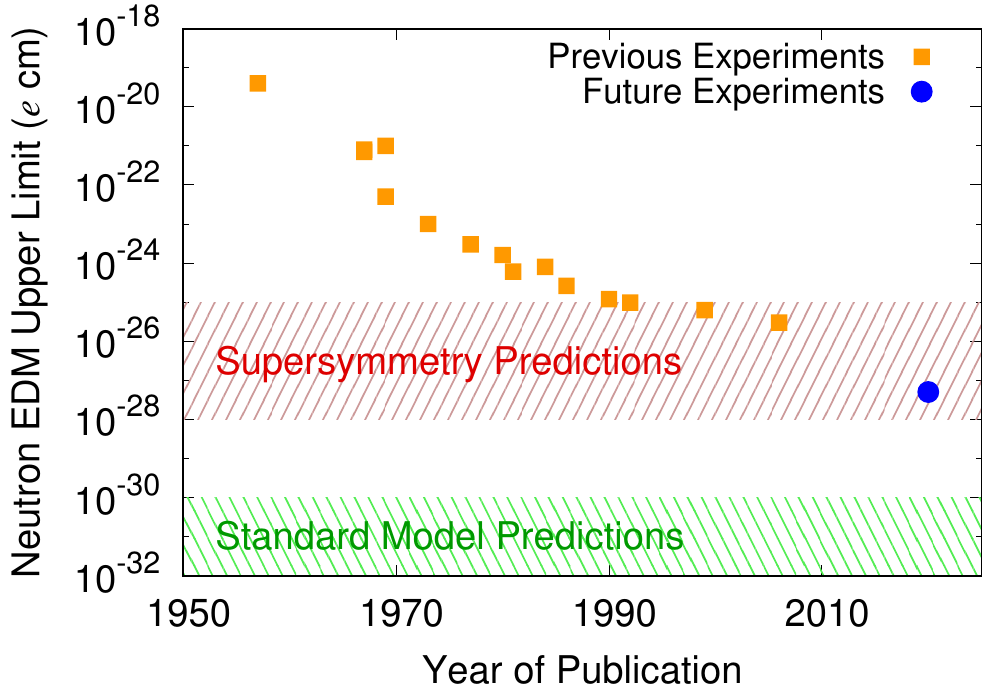}
  \caption{The timeline of the upper bound on the neutron EDM from
    previous and future experiments. The yellow squares are the
    previous bounds, and the blue dot is the sensitivity target of
    next generation experiments. In the standard model, the leading
    contribution to nEDM from the CP violating phase in the CKM matrix
    arises at three-loops, and the expected size of nEDM is more than
    5 orders of magnitude below the current experimental bound, as
    shown by the green shaded area in the plot \cite{Dar:2000tn}. In extensions of the
    standard models, however, nEDM can appear at one-loop with the new
    CP violating interactions. In some of the most popular models,
    such as SUSY, the expected size of nEDM covers the regions of the
    current and future experimental bound, as shown by the red shaded
    area \cite{Pospelov:2005pr, RamseyMusolf:2006vr, Engel:2013lsa}.}
  \label{fig:nedm_exps}
\end{figure}

So far, a nonvanishing neutron EDM (nEDM) has not been observed, and
the current experimental upper bound is $3.0\times 10^{-26} \ecm$~\cite{Afach:2015sja}.
In recent experiments, the nEDM is
measured by the change in the spin precession frequency of
ultracold-neutrons aligned in a magnetic field under a flip in the
direction of a strong background electric
field. Figure~\ref{fig:nedm_exps} shows the history of the
experimental upper bound on the neutron EDM, as well as the target
precision, $\sim 5\times10^{-28}\ecm$ \cite{psi, Picker:2016ygp,
  fermII,sns,CryoEDM}, of proposed experiments.

In the standard model, the leading contribution to nEDM due to the CP
violating phase in the CKM matrix comes from three-loop and higher
order diagrams, and the expected size is more than 5 orders of
magnitude below the current experimental bound \cite{Dar:2000tn}. In extensions of the
standard models, however, nEDM can appear at one-loop due to novel CP
violating interactions. In some of the most popular models, such as
supersymmetric (SUSY) models, the expected size of the nEDM is between
$10^{-25}$--$10^{-28}$$\ecm$ as shown in Figure~\ref{fig:nedm_exps}.
Therefore, a nEDM smaller that $10^{-28}$$\ecm$ will put a serious
constraint on those models \cite{Pospelov:2005pr, RamseyMusolf:2006vr, Engel:2013lsa}.

There are two main outcomes of neutron EDM experiments for BSM
physics. First, a non-zero measurement of nEDM would establish new
sources of CP violation, which is one of Sakharov's three conditions
for weak-scale baryogenesis~\cite{Sakharov:1967dj}. The CP violation
in the standard model is not sufficient to explain observed 
the baryon asymmetry of the universe, and it requires new sources of
CP violation from beyond the standard model (BSM) \cite{Bernreuther:2002uj}. 
Neutron EDM is a good probe for such novel CP violation. Second, most 
BSM theories have additional sources of CP violation. There are many
BSM scenarios predicting a nEDM between $10^{-25}$ -- $10^{-28} \ecm$,
and upcoming experiments will put constraints on
them~\cite{psi,Picker:2016ygp,fermII,sns,CryoEDM}. This requires both the measurement of, or
a bound on, the neutron EDM, and the calculations of the matrix
elements of novel CP violating operators within the neutron states. 
Current non-lattice estimates of the matrix elements have large
uncertainties~\cite{Engel:2013lsa}.  Lattice QCD can provide a first
principle, non-perturbative method for calculating these matrix
elements.

To analyze novel CP violation, we work within the framework of
effective field theories and classify operators by their canonical
dimension~\cite{Pospelov:1999mv}.  At hadronic scale, the effective
Lagrangian for the CP violating interactions at dimension 4, 5 and 6
can be written as
\begin{align}
\mathcal{L}_\text{CPV}^{d=4,5,6} = & 
-\frac{\mathrm{g}^2}{32\pi^2} {\theta} G\tildeG 
-\frac{i}{2}\sum_{q=u,d,s}d_q\bar{q}(\sigma_{\mu\nu} F^{\mu\nu})\gamma_5 q
-\frac{i}{2}\sum_{q=u,d,s}\tilde{d}_q \bar{q}(\sigma_{\mu\nu} G^{\mu\nu})\gamma_5 q
\nonumber \\
&+d_w \frac{\mathrm{g}}{6}f^{abc}G^a_{\mu\nu}\tildeG^{\nu\rho, b}G_\rho^{\mu, c} + \sum_i C_i^{(4q)}O_i^{(4q)}\,,
\label{eq:L_eff}
\end{align}
where $\mathrm{g}$ is the QCD coupling constant, and
$\tildeG^{\mu\nu, b} =
\varepsilon^{\mu\nu\alpha\beta}G^b_{\alpha\beta}/2$.  The first term
on the right hand side (r.h.s) is the dimension-4 $\Theta$-term
already allowed in QCD, and the next two terms are the dimension-5 quark
EDM (qEDM) and quark chromo-EDM (cEDM) terms. There are two types of
dimension-6 terms in the second line in the above equation: the
Weinberg three-gluon operator and various four-quark operators.

If the matrix elements of all the operators are $ O(1)$ and the
anomalous dimensions are small, then the $\mathcal{O}(1/\Lambda_{\rm
  scale}^{D-4})$ suppression implies that the lower mass dimension
operators are more important.  However, it turns out that the current
bound on the nEDM already makes the dimension-4 QCD $\Theta$-term
unnaturally tiny; $\bar{\theta} \le \mathcal{O}(10^{-9} - 10^{-11})$
\cite{Crewther:1979pi, Abada:1990bj, Pospelov:1999mv,
  Mereghetti:2010kp}, where $\bar{\theta}$ is the 
total effective CP violating angle $\bar\theta = \theta + \arg\det M_\textrm{CKM}$ with the 
quark mixing matrix $M_\textrm{CKM}$
. The unexpected smallness of this number is known
as the strong CP problem. One of the popular models explaining the
strong CP problem is the Peccei-Quinn mechanism \cite{Peccei:1977hh,
  Peccei:1977ur}, which requires the existence of the, as yet
unobserved, axion. The two dimension-5 operators typically arise at
the TeV scale as dimension-6 operators involving a Higgs.  At the
hadronic scale, 1~GeV, the Higgs fields are replaced by its vacuum
expectation value, $v_\mathrm{EM}$, and the operator becomes
dimension-5 with the coefficients suppressed by
$v_\mathrm{EM}/M_\mathrm{BSM}^2$. How well these dimensional arguments
work and which terms in Eq.~\eqref{eq:L_eff} make the largest
contribution to nEDM depends on the details of the BSM theory. But,
while the couplings depend on the starting BSM model, the matrix
elements of these operators within the neutron states are model
independent results.  Thus, lattice QCD calculation of the matrix
elements of these operators will play an important role in connecting
the measured neutron EDM and novel CP violation in BSM scenarios,
i.e., knowing the matrix elements along with the bound on (or the
value of) the nEDM, one can put bounds on the couplings and thus on
the parameter space of allowed BSM theories.

In this paper, we review recent progress in Lattice QCD calculations
of the matrix elements of these CP violating operators. We start by
discussing the contribution of the QCD $\Theta$-term to the nEDM and
use it to illustrate the various methods used in the calculations.

\section{QCD \texorpdfstring{$\Theta$}{\text theta}-term} \label{sec:theta}

There are three strategies for the calculation of the contribution of CP violating operators to the 
neutron EDM on the lattice. We illustrate these three approaches 
for the case of the CP-violating $\Theta$-term whose volume integral is the
topological charge $Q_\mathrm{top}$. They are 
the external electric field method
\cite{Aoki:1989rx, Aoki:1990ix, Shintani:2006xr, Shintani:2008nt},
expansion in ${{\theta}}$ \cite{Shintani:2005xg, Berruto:2005hg,
  Shindler:2015aqa, Alexandrou:2015spa, Shintani:2015vsx}, and
simulation with imaginary ${{\theta}}$ \cite{Izubuchi:2008mu, Aoki:2008gv,
  Guo:2015tla}.  

\subsection{External Electric Field Method}

The first method uses an external electric field
\cite{Aoki:1989rx, Aoki:1990ix, Shintani:2006xr, Shintani:2008nt}, in which 
the neutron EDM, $d_N$, is extracted from the energy difference of the neutron
states with spin $\vec{S}$ aligned [anti]parallel to the external electric field
$\vec{\mathcal{E}}$:
\begin{align}
	E^{{\theta}}_{\vec{S}}  - E^{{\theta}}_{-\vec{S}} \approx 2d_N {{\theta}} \vec{S}\cdot\vec{\mathcal{E}}\,.
\end{align}
For a given value of ${{\theta}}$, the effect of the $\Theta$-term is 
included by reweighting the nucleon correlation function with the 
topological charge:
\begin{align}
  \big\langle N\bar{N} \big\rangle_{{\theta}} (\vec{\mathcal{E}}, t)
  = \big\langle N(t) \bar{N}(0) e^{i{{\theta}} Q_{\mathrm{top}}}\big\rangle_{\vec{\mathcal{E}}}\,,
\end{align}
where the expectation value on the r.h.s is evaluated on lattice
configurations generated without the $\Theta$-term.
This method extracts the neutron EDM from a spectral quantity, not the
form factor $F_3$ as discussed next, so the results are not affected by 
the mixing under parity violation problem discussed in Section~\ref{sec:parity}.

\subsection{Expansion in \texorpdfstring{${{\theta}}$}{\text theta}}

Based on the current experimental bound of the neutron EDM and model
studies, we know that the coupling ${\theta}$ is tiny. Hence it can be
treated as a small expansion parameter. Any expectation value in
presence of a small QCD $\Theta$-term can then be written as follows
\cite{Shintani:2005xg, Berruto:2005hg, Shindler:2015aqa,
  Alexandrou:2015spa, Shintani:2015vsx}:
\begin{align}
\big\langle O(x) \big\rangle_{{\theta}}
&= \frac{1}{Z_{{\theta}}} \int d[U, q, \bar{q}] O(x) e^{-S_{QCD}+i{{\theta}} Q_{\mathrm{top}}} \nonumber \\
&= \big\langle O(x) \big\rangle_{{{\theta}}=0} + i {{\theta}} \big\langle O(x) Q_{\mathrm{top}} \big\rangle_{{{\theta}}=0} + O({{\theta}}^2)\,.
\label{eq:theta-exp}
\end{align}
In the second line of the above equation, the QCD $\Theta$-term 
effect is included at the lowest order by measuring 
the correlation of the topological charge with the observable $O$, using 
configurations generated without the $\Theta$-term.

To calculate the nEDM, the operator $O$ is chosen as the matrix
element of the vector current within the neutron states.  Assuming
PT invariance, the matrix element, calculated as a function of $q_\mu$, the
momentum transfered by the vector current, can be decomposed
in terms of Lorentz covariant form factors:
\begin{align}
\langle N V_\mu N \rangle_{{\theta}} = \bar{u}\left[
F_1(q^2)\gamma_\mu
+ i\frac{F_2(q^2)}{2m_N} \sigma_{\mu\nu}q^\nu 
- \frac{F_3(q^2)}{2m_N} \sigma_{\mu\nu} q^\nu \gamma_5\right] u\,.
\label{eq:FF-anly}
\end{align}
Of these, the form factor of interest is $F_3$ at zero-momentum
transfer.  This is extracted by extrapolating $F_3(q^2)$, measured at
finite $q^2$, because the term containing $F_3$ only contributes at
$q^\nu \ne 0$ as can be seen from Eq.~\eqref{eq:FF-anly}. Then the contribution to 
$d_N$ is given by 
\begin{align}
d_N = \lim_{q^2\rightarrow 0}\frac{F_3(q^2)}{2m_N} \,.
\end{align}
This approach, which relies on calculating the form-factor $F_3$ from
nucleon matrix elements, requires a careful handling of the neutron
spinors in a theory with P violation.
Most previous calculations did not correctly account for this
subtlety~\cite{Abramczyk:2017oxr}.  The problem can be corrected 
retroactively, and in Figure~\ref{fig:F3} we show the resulting
reduction in the value reported in recent
calculations~\cite{Alexandrou:2015spa,Shintani:2015vsx}).

\subsection{Imaginary \texorpdfstring{${{\theta}}$}{\text theta} Simulation}

The $\Theta$-term in Euclidean space-time is purely imaginary, and
Monte Carlo simulations including the $\Theta$-term face the sign
problem. One way to circumvent this issue is to make the action
real by taking ${{\theta}}$ to be purely
imaginary~\cite{Izubuchi:2008mu, Aoki:2008gv,Guo:2015tla}.  Assuming
that the theory is analytic around ${{\theta}}=0$, the results can
then be continued to real ${{\theta}}$ for small ${{\theta}}$.
Furthermore, this calculation can be carried out using the axial
anomaly whereby the QCD $\Theta$-term is chirally rotated to the
fermionic term $S_{{\theta}}^q$ as
\begin{align}
{{\theta}} \to  i\tilde{{{\theta}}}, \qquad 
S_{{\theta}}^q = \tilde{{{\theta}}} \frac{m_l m_s}{2m_s + m_l} \sum_x \bar{q}(x) \gamma_5 q(x)\,.
\end{align}
Lattice simulation can then be performed by adding
$S_{{\theta}}^q$ to the original QCD action. In this approach, the
EDM is again extracted from the form factor analysis, so the results
are affected by the mixing induced by parity violation discussed in
Section~\ref{sec:parity}. Again, in Figure~\ref{fig:F3} we show one of
the recent results~\cite{Guo:2015tla} with and without
the correction due to this mixing problem.

\subsection{Variance Reduction using Cluster Decomposition}
In the calculation of neutron EDM induced by the QCD $\Theta$-term
using ${{\theta}}$-expansion or external electric field method, one needs
to calculate correlators reweighted by the topological
charge. Recently, the authors of Ref.~\cite{Liu:2017man} reported an error
reduction technique using cluster decomposition. The main idea is to
calculate the topological charge only in the vicinity of the sink in
the momentum projection summation. Let us consider a neutron two-point
correlator reweighted with the topological charge $Q$:
\begin{align}
  C_R(t) &= \left\langle \sum_x N(\vec{x}, t) \bar{N}(\mathcal{G}, 0) Q_R(x) \right\rangle, \nonumber \\
  Q_R(x) &= \sum_{||x-y||<R} q(y)\,,
  \label{eq:vrcd}
\end{align}
where $\mathcal{G}$ is the source grid. In conventional calculations,
$Q$ is calculated by summing the topological charge density $q(y)$
from all lattice points. In the new formula, however, the topological
charge is calculated only up to a certain radius $R$ from the sink
point under the assumption that correlations due to long-distance
points are negligible.

\begin{figure}[tb]
  \centering
  \includegraphics[width=0.55\textwidth]{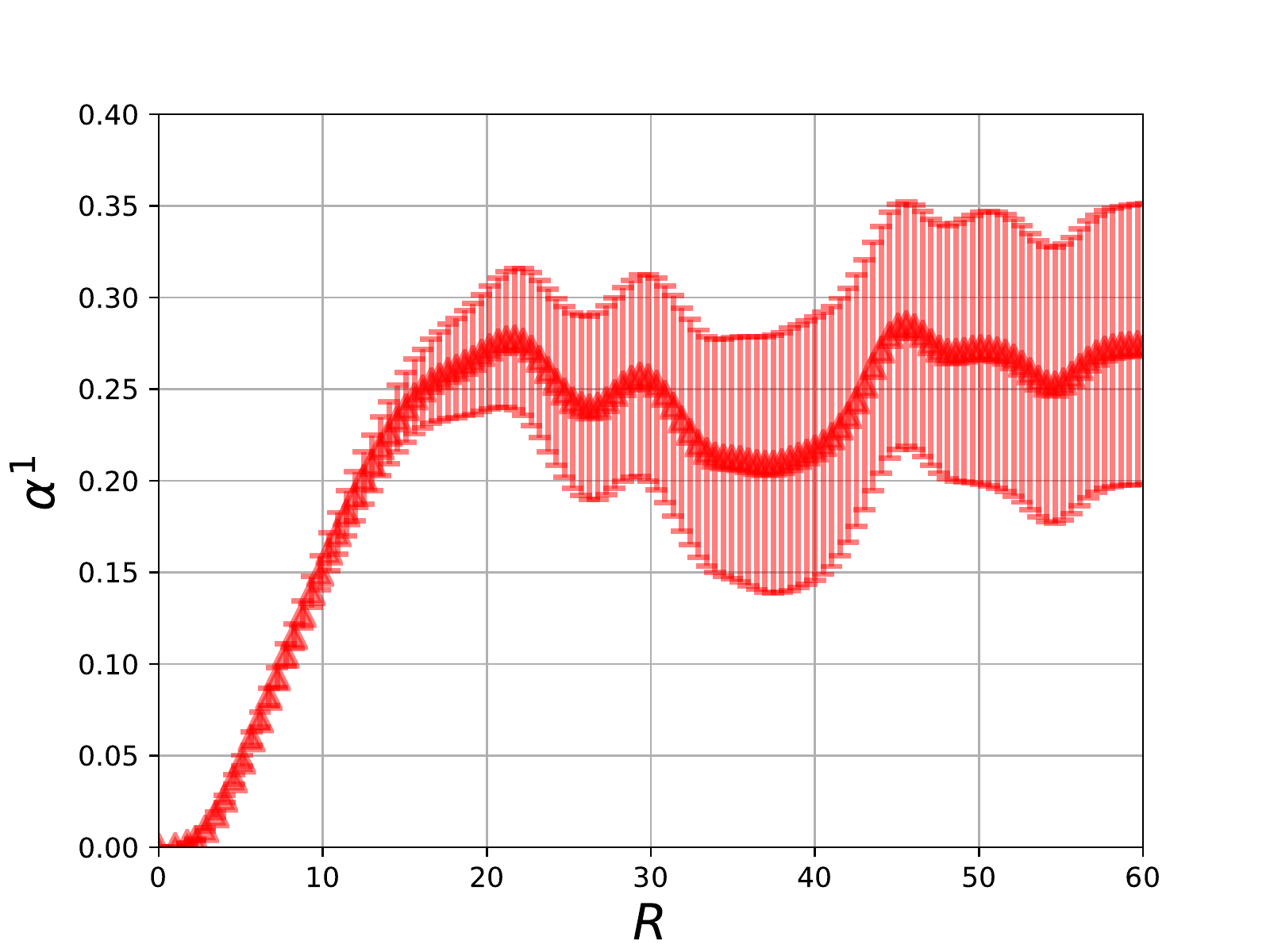}
  \caption{CP violating phase $\alpha$ (denoted as $\alpha^1$ in the figure)
   as a function of the cutoff radius $R$ presented in
    Ref.~\cite{Liu:2017man}. The central value of the $\alpha$ appears
    to saturate after $R$ is larger than about 15 lattice spacings, which 
    corresponds to 1.71~fm in physical unit, while the statistical error
    grows as $R$ is increased.}
  \label{fig:vrcd-alpha}
\end{figure}

Figure~\ref{fig:vrcd-alpha} presents the CP violating phase, labeled
$\alpha^1$ and calculated using the Eq.~\eqref{eq:vrcd}, as a function
of the cutoff radius $R$. Data show that the mean value of $\alpha^1$
saturates at some reasonably small value of $R$. Including the
contributions of the topological density from larger $R$ only
increases the statistical noise. Considering that the mean value still
exhibits significant fluctuations, a more detailed cost-benefit
analysis of the approximation is required.

\section{Extracting \texorpdfstring{$F_3$}{F\string_3} in a Theory with Parity Violation}
\label{sec:parity}

In a theory with P and CP symmetry, the spinor $u$ of the neutron state satisfies the following Dirac equation:
\begin{align}
\left(i p_\mu \gamma_\mu + m\right)u = 0\,, 
\end{align}
with $\gamma_4$ the parity operator:
$u_{\vec{p}} \rightarrow \gamma_4 u_{-\vec{p}}$ under P. When CP, but
not PT, is violated, the Dirac equation is only modified by a phase
factor to
\begin{align}
\left(i p_\mu \gamma_\mu + m e^{-2i\alpha\gamma_5}\right)\tilde{u} = 0\,,
\end{align}
where the spinor that solves the modified equation is defined to be
$\tilde{u}$. Furthermore, for this equation, $\gamma_4$ is no longer the parity
operator, rather under parity $\tilde{u}_{\vec{p}} \rightarrow e^{-2i\alpha\gamma_5} \gamma_4
\tilde{u}_{-\vec{p}}$. The two solutions $u$ and $\tilde{u}$ are related as 
\begin{align}
\tilde{u} = e^{i\alpha\gamma_5}u, \qquad \bar{\tilde{u}} = \bar{u}e^{i\alpha\gamma_5}.
\label{eq:phase}
\end{align}

When calculating the form factors $F_2$ and $F_3$, it is important to
enforce that $F_2$ is the parity-even magnetic form factor while $F_3$
is the P and CP odd electric form factor. There are two ways to
enforce these symmetry requirements.  The first is to properly include
the phase defined in Eq.~\eqref{eq:phase} in the definition of the
matrix element. In that case the decomposition into the form factors
is the same as in a theory with CP symmetry with unmodified spinors
$u$. Alternatively, one can calculate the standard matrix element, and
then undo the mixing between $F_2$ and $F_3$ due to the non-standard
parity transformation with phase $\alpha$. Then, the two form factors
$F_2$ and $F_3$ are given in terms of $\tilde{F}_2$ and $\tilde{F}_3$
as in Eq.~\eqref{eq:parity_mix}.

In both cases, the phase $\alpha$ has to be extracted from the nucleon
2-point function.  In this extraction, it is important to note that
the phase $\alpha$ is state dependent. In Section~\ref{sec:cedm}, we
show that the $\alpha$ corresponding to the ground state is given by
the long-time behavior of the nucleon 2-point function, i.e., when the
ground state dominates.

In the first approach, one includes the phase in the calculation of the n-point functions: 
\begin{align}
e^{-i\alpha\gamma_5} \langle \tilde{N} V_\mu \bar{\tilde{N}} \rangle e^{-i\alpha\gamma_5}
= \langle N V_\mu \bar{N} \rangle , \qquad
e^{-i\alpha\gamma_5} \langle \tilde{N} \bar{\tilde{N}} \rangle e^{-i\alpha\gamma_5}
=  \langle N\bar{N}\rangle \nonumber  \\
\Longrightarrow \quad
\bar{u} \left(F_1(q^2)\gamma_\mu
+ i\frac{[\gamma_\mu,\gamma_\nu]}{2}q_\nu \frac{F_2(q^2)}{2m_N}
- \frac{[\gamma_\mu,\gamma_\nu]}{2} q_\nu \gamma_5 \frac{F_3(q^2)}{2m_N}\right) u.
\end{align}
The resulting $F_3(0)/2M_N$ is the desired contribution to the nEDM.

In the second approach, one calculates 
\begin{align}
\langle \tilde{N} V_\mu \tilde{\bar{N}} \rangle, \quad
\langle \tilde{N} \tilde{\bar{N}} \rangle
\quad \Longrightarrow \quad
\bar{\tilde u}\left(\tilde{F}_1(q^2)\gamma_\mu
+ i\frac{[\gamma_\mu,\gamma_\nu]}{2}q_\nu \frac{\tilde{F}_2(q^2)}{2m_N}
- \frac{[\gamma_\mu,\gamma_\nu]}{2} q_\nu \gamma_5 \frac{\tilde{F}_3(q^2)}{2m_N}\right)\tilde u\,.
\end{align}
and one has to extract $F_3$ from the following mixing structure: 
\begin{align}
\begin{split}
  F_2 = \cos(2\alpha) \tilde{F}_2 - \sin(2\alpha)\tilde{F}_3\,, \\
  F_3 = \sin(2\alpha) \tilde{F}_2 + \cos(2\alpha)\tilde{F}_3\,.
\end{split}
\label{eq:parity_mix}
\end{align}
The two approaches are equivalent. 

Unfortunately, as pointed out in Ref.~\cite{Abramczyk:2017oxr}, all
calculations based on the $F_3$ extraction prior to their work, 
starting with the 2005 work in Ref.~\cite{Shintani:2005xg}, used the 
second approach but did not carry out the rotation defined in
Eq.~\eqref{eq:parity_mix} to get the $F_3$. Their quoted results are,
therefore, for $\tilde{F}_3$. The authors of
Ref.~\cite{Abramczyk:2017oxr} show that correcting for this omission
reduces the value of an already hard to measure $F_3$ by about a
factor of ten.  Thus, all previous estimates of the contribution of the 
$\Theta$-term and the cEDM based on evaluating $F_3$ need to be revised.

\begin{figure}[thb]
  \centering
  \includegraphics[width=0.6\textwidth]{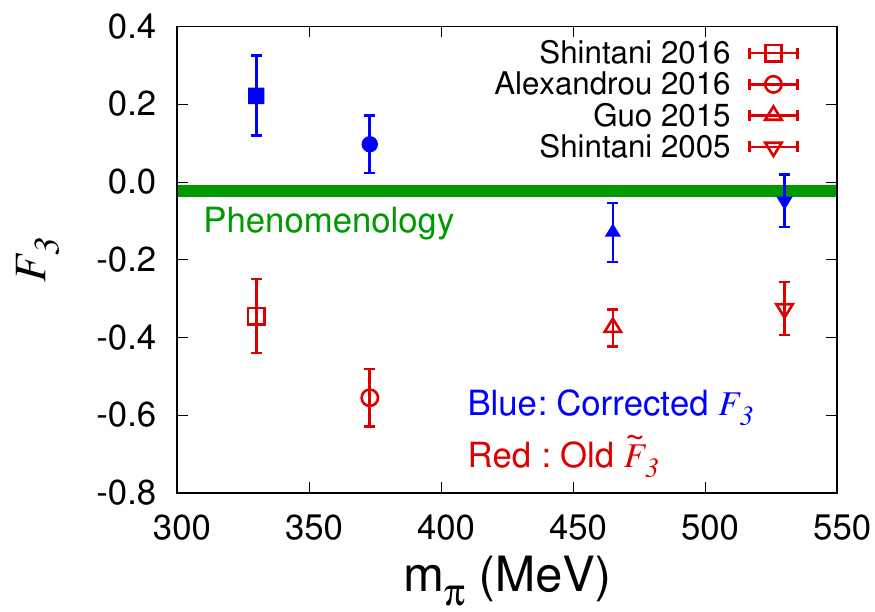}
  \caption{Recent lattice results of $F_3$ induced by the QCD
    $\Theta$-term calculated with $M_\pi < 550$~MeV. Red points are
    the original results, which give $\tilde{F}_3$, and the blue
    points are the corrected results. Original results are obtained
    from Refs. \cite{Shintani:2015vsx} (labeled Shintani 2016),
    \cite{Alexandrou:2015spa} (Alexandrou 2016),
    \cite{Guo:2015tla} (Guo 2015), and \cite{Shintani:2005xg}
    (Shintani 2005), and the corrected results are obtained
    from Refs.~\cite{Abramczyk:2017oxr, Shintani:priv_com}. As noted
    in Ref.~\cite{Abramczyk:2017oxr}, some assumptions are made to
    calculate the corrected $F_3$ numbers as some of the original
    papers do not contain full information needed for the parity
    induced rotation Eq.~\protect\eqref{eq:parity_mix}, so the $F_3$
    results presented in this plot may not be precise.}
  \label{fig:F3}
\end{figure}

Figure~\ref{fig:F3} shows some of the recent lattice results of $F_3$
induced by the QCD $\Theta$-term, before and after the parity mixing
correction. The interesting point is that all corrected numbers are
close to zero. Previous phenomenological estimates were an order of
magnitude smaller than lattice QCD results. This tension may
disappear after a proper analysis of the phase $\alpha$ introduced in
the neutron states due to the breaking of parity. Further discussion
of the mixing between $F_2$ and $F_3$ when parity is violated can be
found in Ref.~\cite{Abramczyk:2017oxr}.

\section{quark EDM}\label{sec:qedm}

The electromagnetic current, $V_\mu \equiv \delta \cal{L}/\delta
A_\mu$, gets an additional contribution when the quark EDM operator is
added to $\cal{L}_{\rm SM}$. Thus, while the standard conserved vector
current interacts with a quark with charge $q_f$, this additional
contribution, which is is just the flavor diagonal tensor operator for
each quark flavor, couples with strength $d_f^\gamma g_T^f$, where
$d_f^\gamma $ are the BSM model dependent CP violating couplings and
$g_T^u$, $g_T^s$, $g_T^s$, $\ldots$ are the tensor charges, i.e.,
matrix elements of the quark bilinear tensor operator within the
nucleon states. Furthermore, the effects analogous to the mixing
discussed in Section~\ref{sec:parity} are suppressed by powers of the
electromagnetic coupling \(\alpha_{\rm EM}\sim1/137\). Thus, the
leading contribution of the EDM of the quarks to the nEDM is given by
the flavor diagonal tensor charges: %
\begin{align}
&\big\langle N \vert \bar{q} \sigma_{\mu\nu} q \vert N \big\rangle 
 =g_T^q \bar{u}_N \sigma_{\mu\nu} u_N\,, \nonumber \\
&d_N = d_u g^u_T + d_d g^d_T +d_s g^s_T\,.
\label{eq:qedm}
\end{align}
In many models, such as the supersymmetric models, the quark EDMs are
proportional to the corresponding quark masses ($d_q \propto m_q$),
because in these theories all connections between left and
right-handed quarks are mediated by a common set of Yukawa
interactions. Since the strange quark is much heavier than the u and the  d
quarks, the strange quark contribution gets enhanced,
by $m_s/m_d \approx 20$. On the other hand $g_T^s \ll g_T^l$. Since
the contribution is the product of the two, it is, therefore,
important to determine the strange quark, and perhaps even the charm
quark, tensor charge precisely.

\begin{figure}[tb]
  \centering
  \includegraphics[width=0.45\textwidth]{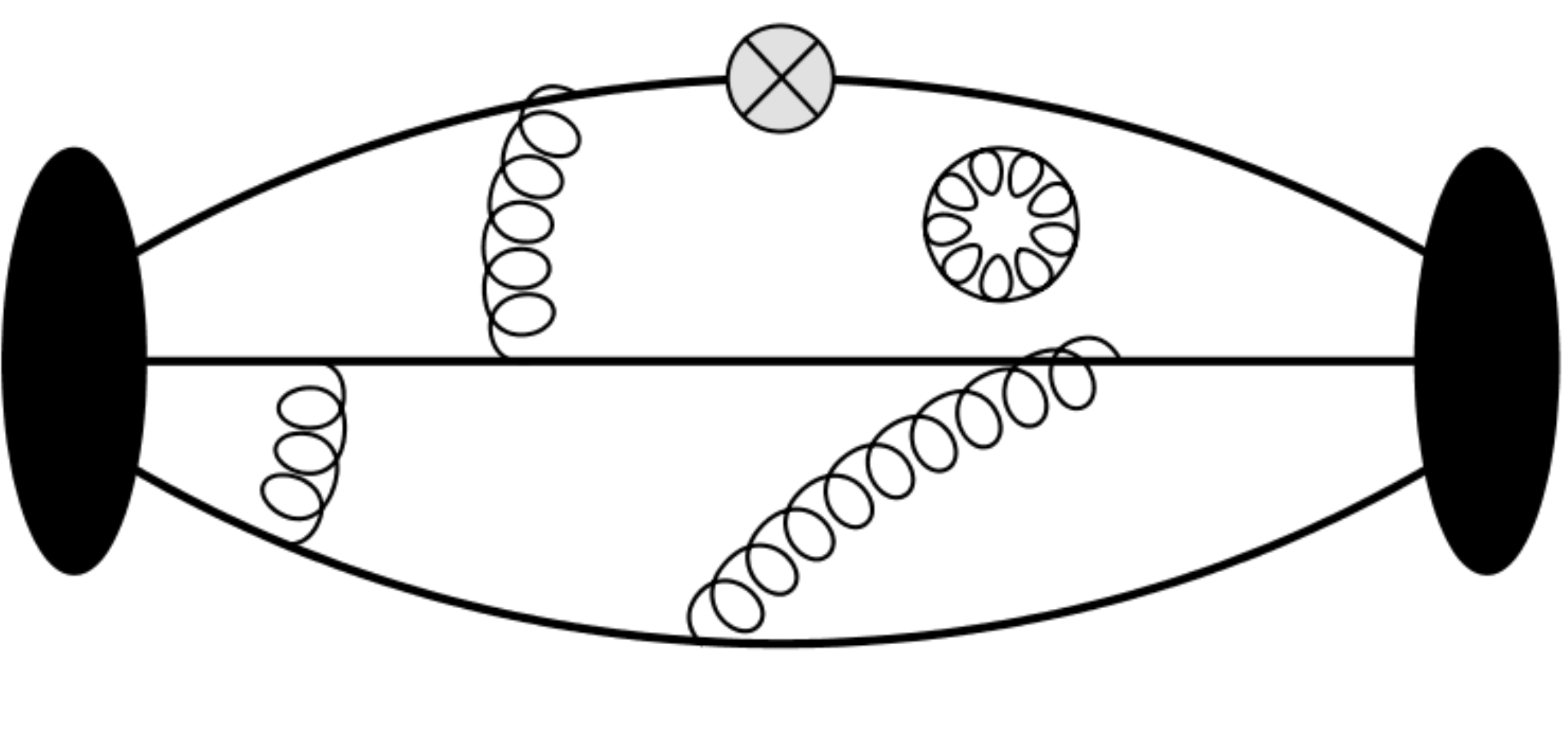}\qquad
  \includegraphics[width=0.45\textwidth]{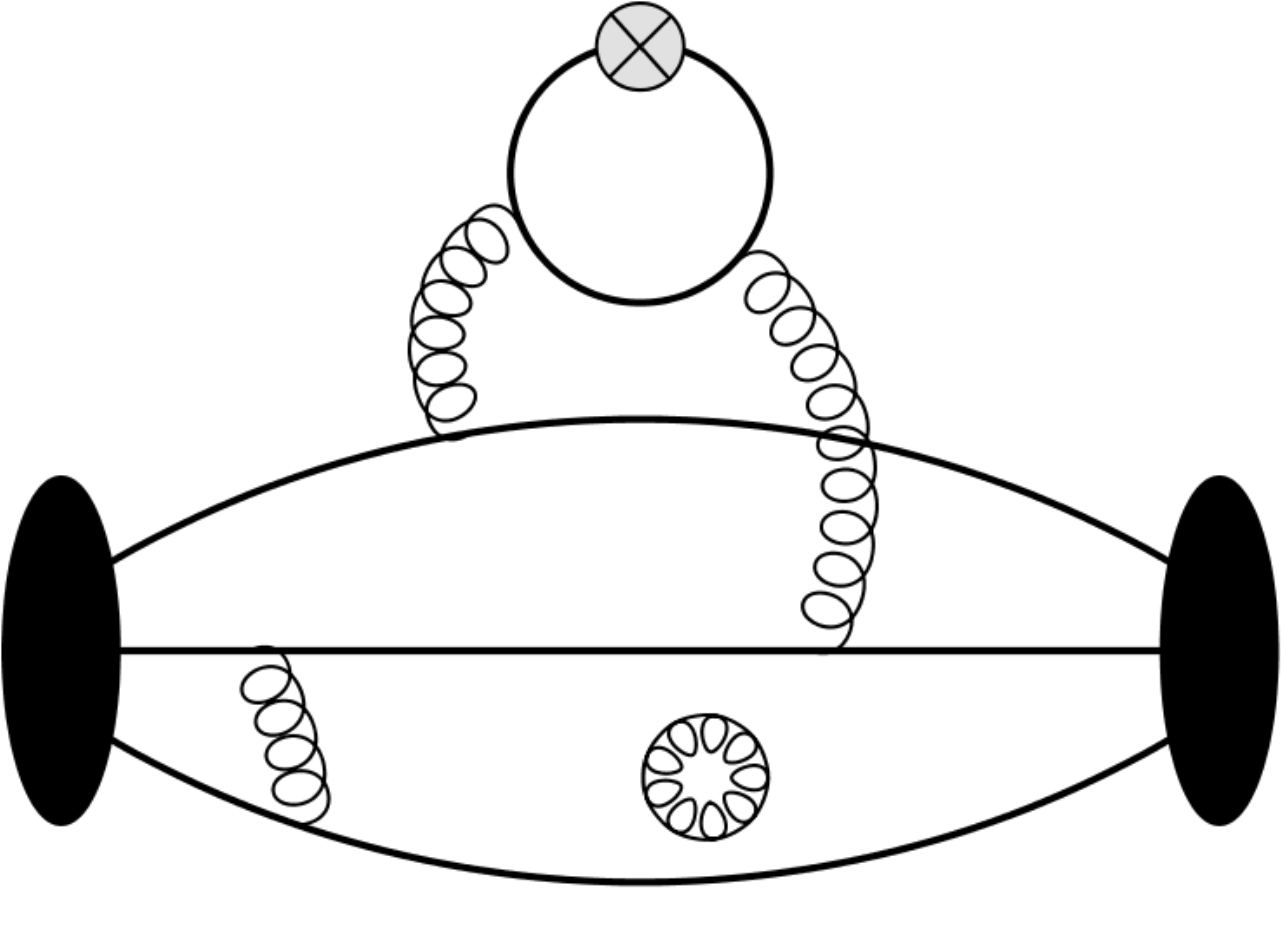}
  \caption{Quark-line connected (left) and disconnected (right) diagrams contribute to the quark EDM. Black ellipses are the neutrons, and the cross is the quark bilinear operator with tensor structure.}
  \label{fig:conn_disc}
\end{figure}

There are two classes of diagrams in the calculation of flavor
diagonal tensor charges: quark-line connected and the quark-line
disconnected diagrams as illustrated in Figure~\ref{fig:conn_disc}. In
case of the strange (charm) quark tensor charge, only the disconnected
diagram with strange (charm) quark loop contributes. This is the basis
of the expectation that $g_T^s \ll g_T^l$.

\begin{figure}[tb]
  \centering
  \includegraphics[width=0.45\textwidth]{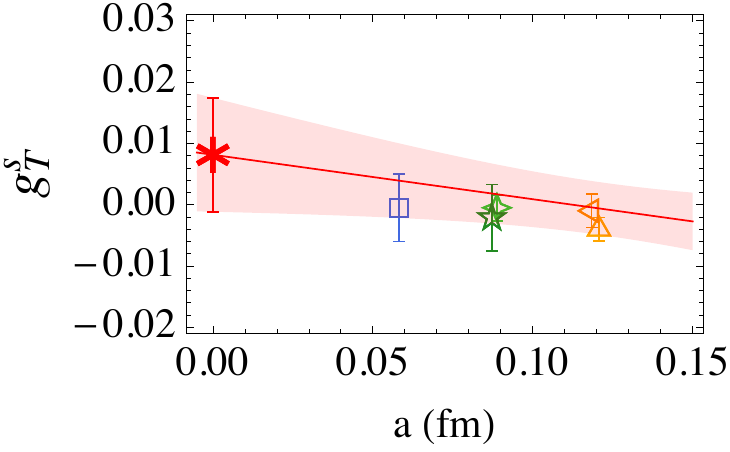}\qquad
  \includegraphics[width=0.45\textwidth]{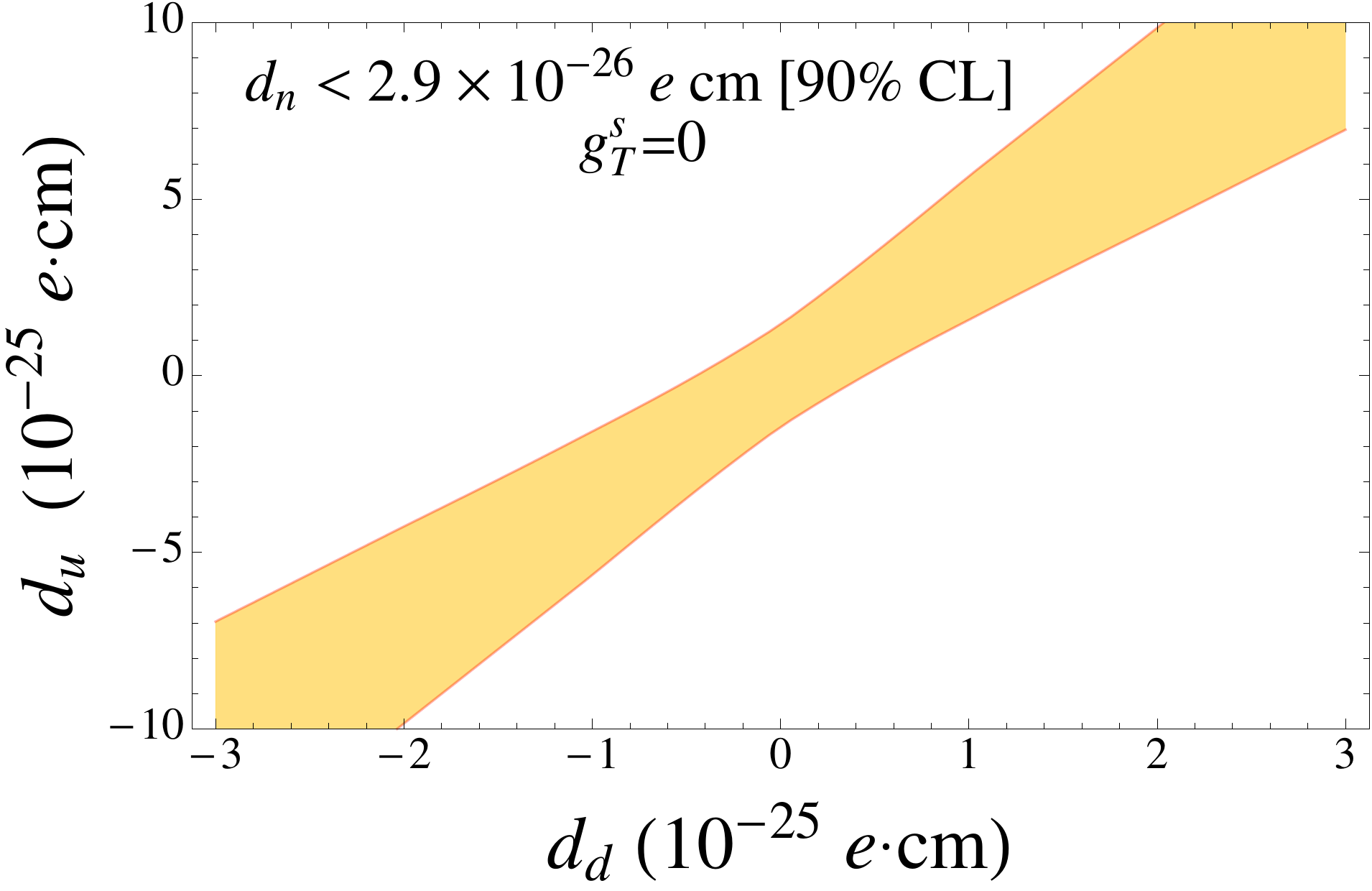}
  \caption{(Left) Extrapolation of $g_T^s$ to the continuum limit. The
    fit is performed simultaneously in $a$ and $M_\pi$ but only the
    projection to the $a$-plane with data extrapolated to the physical
    $M_\pi$, is presented.  (Right) Bounds on $d_u$ and $d_d$ are
    obtained using the formula given in Eq.~\eqref{eq:qedm}, the
    tensor charges given in Eq.~\protect\eqref{eq:tensor-charge} and
    the experimental bounds $|d_N| < 2.9 \times 10^{-26}\ecm$. Since
    the strange quark tensor charge is consistent with zero within
    statistics, it imposes no constraint on $d_s$. Both figures are taken
    from Ref.~\cite{Bhattacharya:2015esa}.}
  \label{fig:qedm}
\end{figure}

In Refs.~\cite{Bhattacharya:2015esa, Bhattacharya:2015wna}, the tensor
charges were calculated using clover fermions on the HISQ lattices, and
extrapolated to continuum and physical pion mass limit, as illustrated
in Figure~\ref{fig:qedm} (left). The disconnected diagrams were
estimated using a stochastic method, and it turned out that the
disconnected diagram contribution to the tensor charges is very
small. Their results are 
\begin{align}
g_T^u = -0.23(3), \quad g_T^d = 0.79(7), \quad g_T^s = 0.008(9)\,.
\label{eq:tensor-charge}
\end{align}

There are BSM models in which the quark EDMs are the dominant sources
of CP violation at low energy. In these scenarios, constraints can be
placed on the couplings of the EDM of each quark flavor using the
experimental bound on the neutron EDM and the lattice results for the
tensor charges. Neglecting all other sources of CP violation,
figure~\ref{fig:qedm} (right) shows the allowed region of $d_u$ and
$d_d$ using the lattice results, Eq.~\eqref{eq:tensor-charge},
described in Ref.~\cite{Bhattacharya:2015esa, Bhattacharya:2015wna}.
Of all the CP violating operators listed in Eq.~\eqref{eq:L_eff}, the
calculation of the quark EDM, including renormalization, is
theoretically well-established and the results in
Eq.~\eqref{eq:tensor-charge} for $g_T^u$ and $g_T^d$ indicate that
these are already available at the $15\%$ level. Further improvement in
precision will occur as higher precision data are generated in future
calculations.

\section{Quark Chromo-EDM}
\label{sec:cedm}

The last dimension-5 term is the quark chromo-electric dipole moment
(cEDM) operator. The modification to the action in presence of the CP violating cEDM term is:
\begin{align}
  S = S_{QCD} - \frac{i}{2} \int d^4x \ \tilde{d}_q
    \bar{q} (\sigma \cdot G) \gamma_5 q\,,
\end{align}
where $\tilde{d}_q$ are again BSM couplings evolved to the hadronic
scale for each quark flavor.  The lattice calculation consists of
evaluating the matrix element of the insertion of the product of the
electromagnetic current \(V_\mu\) and the $\bar{q} (\sigma \cdot G)
\gamma_5 q$ operator within the neutron states. Then, the contribution
of cEDM operators to the nEDM, ignoring the complicated operator
mixing problem under renormalization discussed in
Section~\ref{sec:cEDMrenorm} below, is obtained from $\sum_q
\tilde{d}_q \langle N| V_\mu \bar{q} (\sigma \cdot G) \gamma_5 q | N
\rangle $.

Lattice QCD studies of the cEDM operator have started only recently,
and three methods are being explored: an expansion in $\tilde{d}_q$
\cite{Abramczyk:2017oxr}, external electric field method
\cite{Abramczyk:2017oxr}, and the Schwinger source method
\cite{Bhattacharya:2016oqm, Bhattacharya:2016rrc}.

\subsection{Expansion in  \texorpdfstring{$\tilde{d}_q$}{d\string~\string_q}}

The expansion in $\tilde{d}_q$ method proceeds analogously to the
expansion in ${{\theta}}$ in the QCD $\Theta$-term calculation. For small
BSM cEDM couplings, $\tilde{d}_q$, the neutron matrix element of the
electromagnetic current can be written in terms of the expectation
values evaluated on the standard CP even lattices,
\begin{align}
  \big\langle N V_\mu \bar{N} \big\rangle_{CPV}
  & = \big\langle N V_\mu \bar{N} \big\rangle
   + \tilde{d}_q \Big\langle N V_\mu \bar{N} \cdot 
     \sum_x O_{\mathrm cEDM}(x) \Big\rangle
   + \mathcal{O}(\tilde{d}_q^2)\,, \\
   O_{\mathrm cEDM} &= \frac{i}{2} \bar{q} (\sigma \cdot G) \gamma_5 q\,.
\end{align}
Their contribution to the neutron EDM is extracted from the CP
violating form factor, $F_3$ as defined in Eq.~\eqref{eq:FF-anly}. The
calculation is challenging because it requires calculating the expectation value
of four-point functions because one has to insert both the vector current
and the cEDM operator between the neutron source and sink,
with the cEDM operator defined as a sum over all space-time points.

\begin{figure}[tb]
  \centering
  \includegraphics[width=0.95\textwidth]{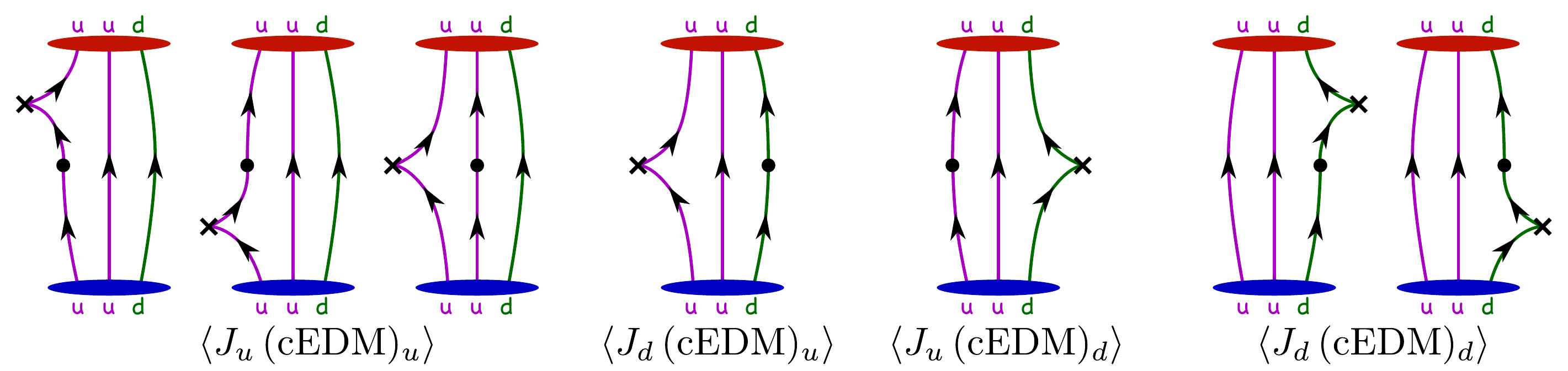}\\
  \includegraphics[width=0.55\textwidth]{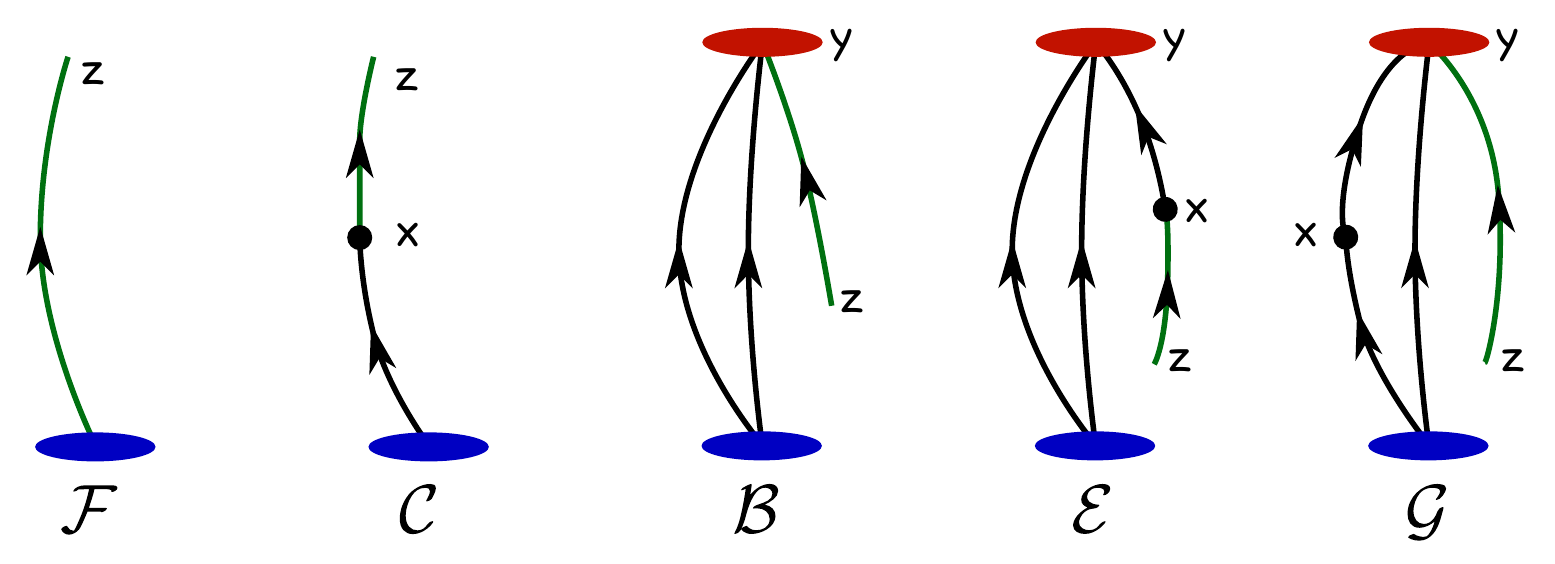}
  \caption{(Top) Connected diagrams needed for the calculation of the cEDM operator. Crosses are the cEDM operator insertion, and dots are the vector current insertion.
  (Bottom) Propagators needed for the quark-line connected four-point correlators. In this figure, the electromagnetic current is denoted as $J$.
  Figures are taken from Ref.~\cite{Abramczyk:2017oxr}.}
  \label{fig:cedm_4p}
\end{figure}

Figure~\ref{fig:cedm_4p} (top) shows all the required quark-line
connected diagrams. These four-point
correlators are constructed using five types of propagators used
as building blocks that are shown in the bottom panel of
Figure~\ref{fig:cedm_4p} (bottom). $\mathcal{F}$ and $\mathcal{B}$ are
the usual forward and backward propagators. $\mathcal{C}$ is the
cEDM-sequential propagator. It is constructed by starting with the
regular quark propagator and inserting the chromo-EDM operator at all
lattice sinks points. This is then used as the source for calculating
the chromo-EDM inserted propagator. 
 
\begin{figure}[tb]
  \centering
  \includegraphics[width=0.45\textwidth]{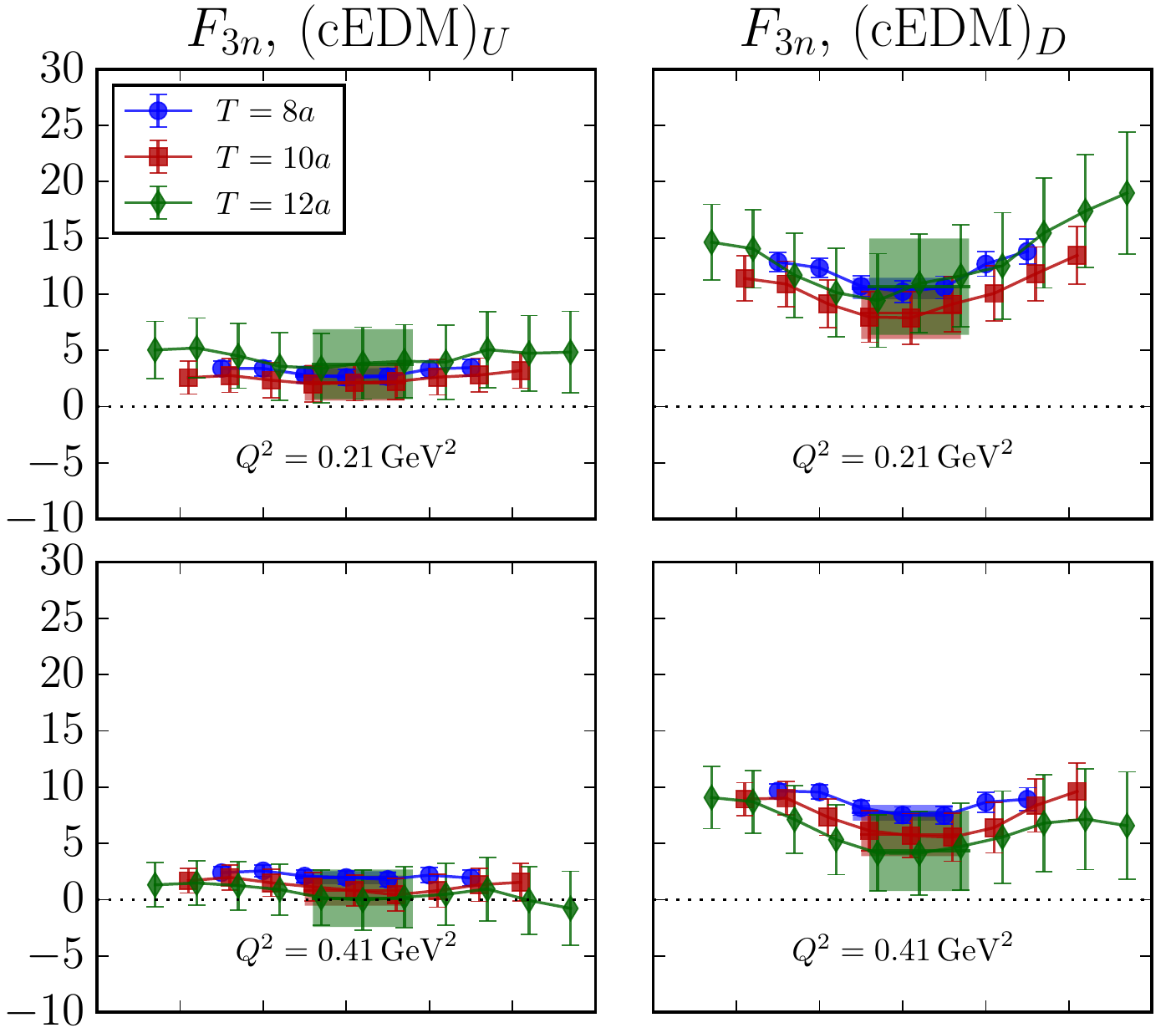}\qquad
  \includegraphics[width=0.45\textwidth]{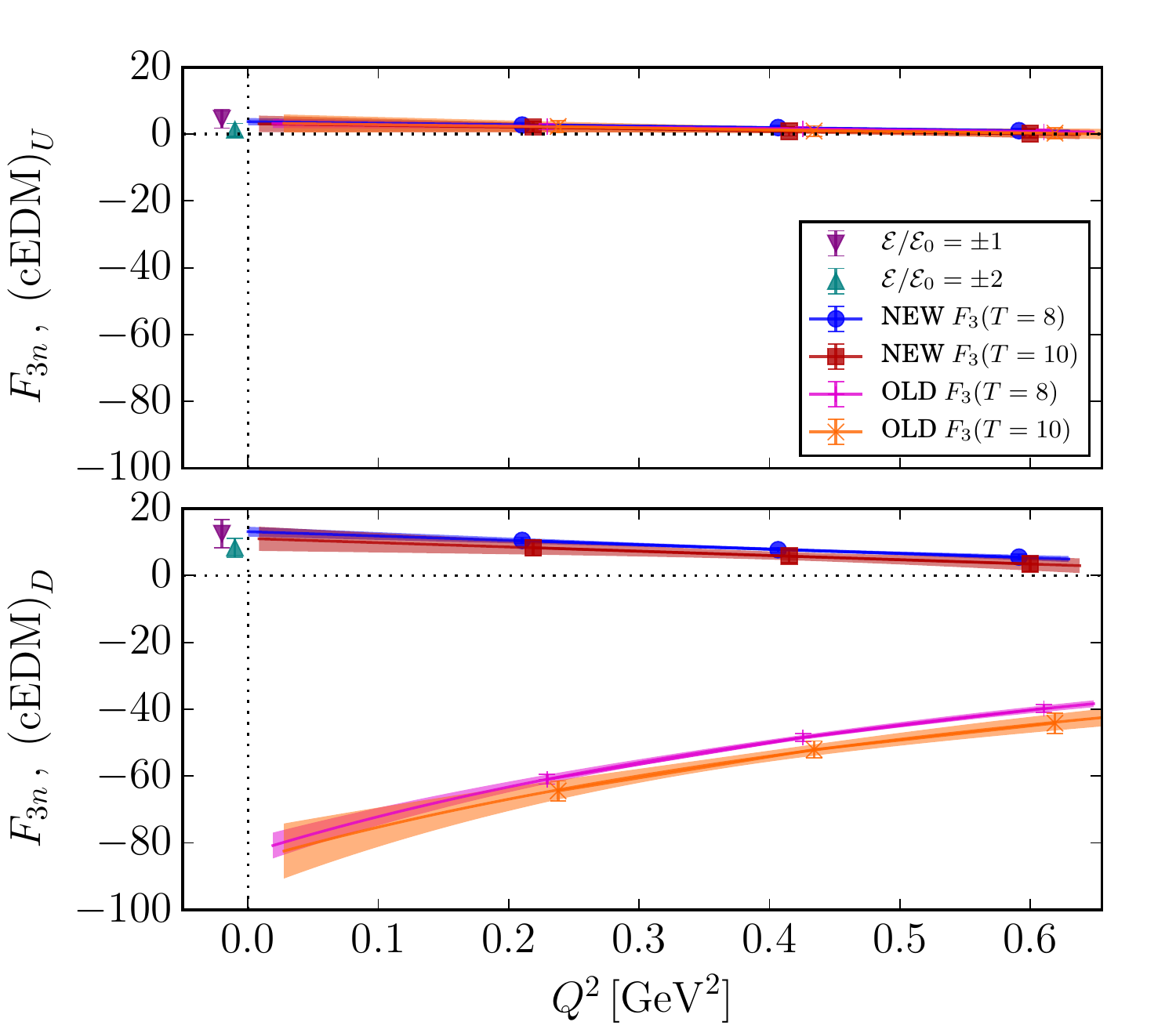}
  \caption{(Left) Plateau plots for $F_3$ induced by the cEDM operator
    calculated by expanding in $\tilde{d}_q$. Simulations were
    performed on $24^3\times 64$, $a=0.11$~fm and $M_\pi=340$~MeV DWF
    lattices.  (Right) Results for cEDM contribution to $F_3$
    extrapolated to $Q^2=0$ from the $\tilde{d}_q$-expansion results
    compared with the $F_3$ obtained by using the external electric
    field method. OLD and NEW $F_3$ label the results before and after
    fixing the parity mixing rotation given in
    \protect\eqref{eq:parity_mix}. Same lattices used in (left) plot
    are used for this
    calculation. $\mathcal{E}_0=0.039~\mathrm{GeV}^2$.  Note that these 
    results are not renormalized.  Figures are taken from
    Ref.~\cite{Abramczyk:2017oxr}.}
  \label{fig:cedm1}
\end{figure} 

Results from Ref.~\cite{Abramczyk:2017oxr} for the signal in $F_3$ and
the source-sink separation dependence of the signal are shown in
Figure~\ref{fig:cedm1} (left). $F_3$ has larger value when the cEDM is
inserted in the  d-quarks of the neutron. In Figure~\ref{fig:cedm1}
(right), the results are extrapolated to zero-momentum to obtain the
contribution $F_3(0)$ as needed for the neutron EDM.

\subsection{External Electric Field Method}

The chromo-EDM contribution to the neutron EDM also can be calculated
using the external electric field. Similarly to the QCD $\Theta$-term,
the neutron EDM is extracted from the energy difference of nucleon
states under spin flip in presence of a uniform external electric
field $i\mathcal{E}$:
\begin{align}
	E^\mathrm{cEDM}_{\vec{S}}  - E^\mathrm{cEDM}_{-\vec{S}} \approx 2d_N \vec{S}\cdot i\vec{\mathcal{E}}\,,
\end{align}
where the neutron correlators in the presence of the cEDM term are obtained by reweighting,
\begin{align}
  \big\langle N\bar{N} \big\rangle_\mathrm{cEDM} (i\vec{\mathcal{E}}, t)
  = \big\langle N(t) \bar{N}(0) O_\mathrm{cEDM} \big\rangle_{i\vec{\mathcal{E}}} \,.
\end{align}
This method needs only two- and three-point functions of the
neutron, the latter with the insertion of the cEDM bilinear operator. In
Ref.~\cite{Abramczyk:2017oxr}, $F_3$ is extracted using the external
electric field method on the same ensemble of lattices used for the calculation
using the $\tilde{d}_q$-expansion method, and the results are presented in
Figure~\ref{fig:cedm1} (right). The signal is poorer in the external
electric field method compared to those in expansion in $\tilde{d}_q$ method, 
but the results of the two methods are
consistent. Note that the external electric field method is not
affected by the parity mixing problem described in
Section~\ref{sec:parity}, while the $\tilde{d}_q$-expansion method
is. Therefore, the consistency of the results from the two methods, 
after taking care of the parity mixing rotation given in
\protect\eqref{eq:parity_mix}, suggests that both methods are 
yielding a reliable signal. 

\subsection{Schwinger Source Method}
\label{sec:SSM}

Another method to calculate the contribution of the cEDM operator to
the neutron EDM is the Schwinger source method
\cite{Bhattacharya:2016oqm, Bhattacharya:2016rrc}. Noting that the
cEDM operator is a quark bilinear, $i \bar{q} (\sigma \cdot G)\gamma_5
q$, one can add it to the QCD fermion action:
\begin{align}
  D_\mathrm{clov} \longrightarrow 
  D_\mathrm{clov} + i\varepsilon \sigma^{\mu\nu} \gamma_5 G_{\mu\nu}\,,
\label{eq:LcEDM}
\end{align}
where $D_\mathrm{clov}$ is the Dirac operator for the Wilson-clover
action, and $\varepsilon$ is a small control parameter the dependence
on which will be removed by taking a derivative with respect to it at
the end of the calculation. Because cEDM is a quark bilinear, the
integration over the fermion degrees of freedom in the path integral
can still be carried out as before.  In the case of the clover
fermions, the addition of the cEDM operator is equivalent to shifting
the coefficient of the dimension-5 clover term by
$i\varepsilon\gamma_5$:
\begin{align}
  c_\mathrm{sw} \sigma^{\mu\nu} G_{\mu\nu} \longrightarrow 
  \sigma^{\mu\nu} (c_\mathrm{sw} + i\varepsilon \gamma_5) G_{\mu\nu}\,.
\end{align}
In this approach, the insertion of the cEDM term is carried out by
calculating valence quark propagators including this modified clover term,
and using this propagator in all diagrams requiring a cEDM
insertion. Having taken care of the cEDM term, the calculation of the
original 4-point functions reduces to three-point functions, 
with an insertion of the vector current within the neutron state, made
up of propagators with and without cEDM insertion. 

\begin{figure}[tb]
  \centering
  \includegraphics[width=0.98\textwidth]{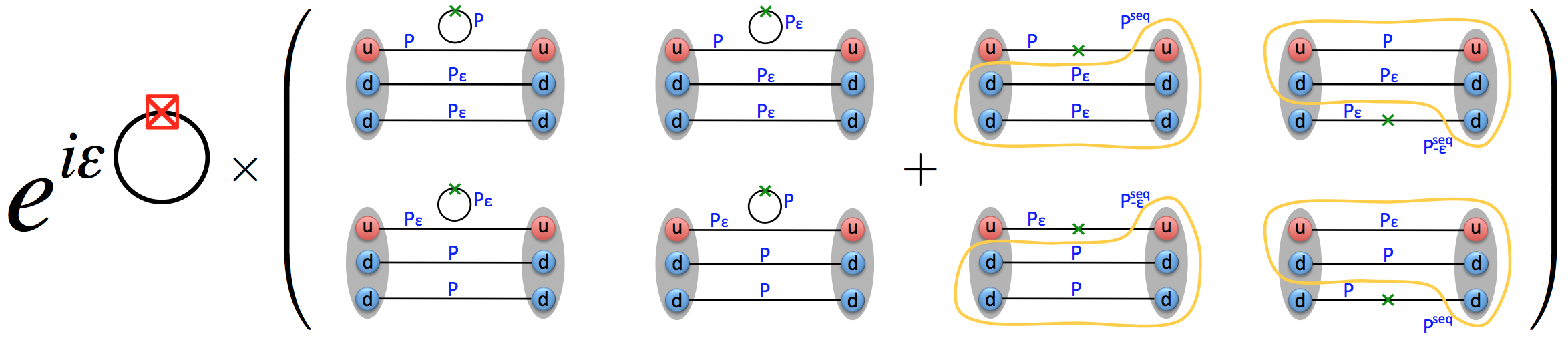}
  \caption{Quark line diagrams needed for the Swinger source
    method. The reweighting factor (left), the quark-line disconnected
    (middle four), and the quark-line connected (right four)
    diagrams. Red crossed box is the cEDM operator insertion, $P$ is
    the regular quark propagator without the cEDM operator insertion,
    and $P_\varepsilon$ is the quark propagator with the cEDM operator
    insertion. An identical calculation has to be done with the cEDM
    operator replaced by the $\gamma_5$ operator in order to define a
    finite renormalized quantity.}
  \label{fig:cedm-swinger-diag}
\end{figure}

To perform this calculation on ensembles generated with just the standard QCD action, however, 
requires taking into account the change in the Boltzmann factor under
the addition of the cEDM term to the action as shown in Eq.~\eqref{eq:LcEDM}. This,
a priori, non-unitary formulation between sea and valence quarks can be
accounted for at leading order in $\varepsilon$ by reweighting each
configuration by ratio of the fermion determinant with and without the
cEDM term:
\begin{align}
\frac{\det\Big( D_\mathrm{clov} + i\varepsilon \sigma^{\mu\nu} \gamma_5 G_{\mu\nu} \Big)}
     {\det\Big( D_\mathrm{clov} \Big)}
\approx
\exp\Big[ i\varepsilon \mathrm{Tr}\big( \sigma^{\mu\nu} \gamma_5 G_{\mu\nu} D_\mathrm{clov}^{-1} \big) \Big]\,.
\end{align}
Diagrammatically, all the terms that need to be calculated in this
Schwinger source method are shown in
Figure~\ref{fig:cedm-swinger-diag}. There are three classes of
diagrams at leading order in $\varepsilon$: the reweighting factor for
each configuration, and the quark-line connected and disconnected
diagrams on that configuration.  

\begin{figure}[tb]
  \centering
  \includegraphics[width=0.31\textwidth]{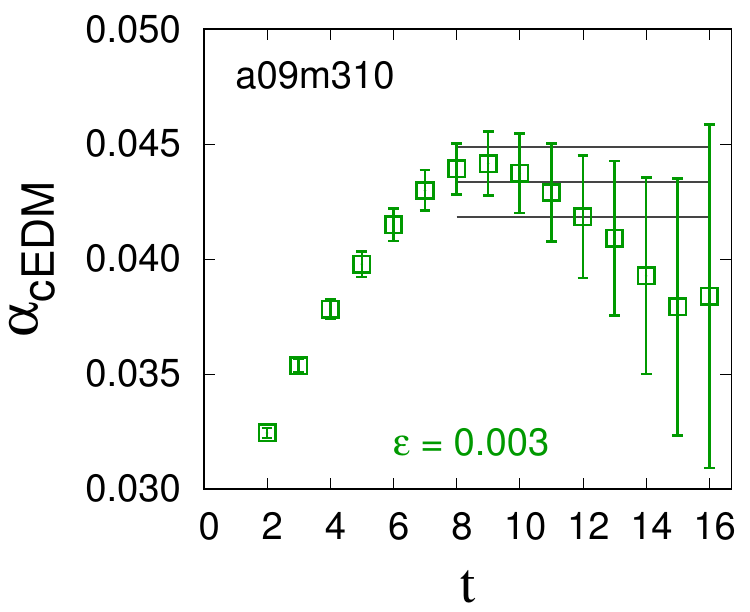}  \quad
  \includegraphics[width=0.31\textwidth]{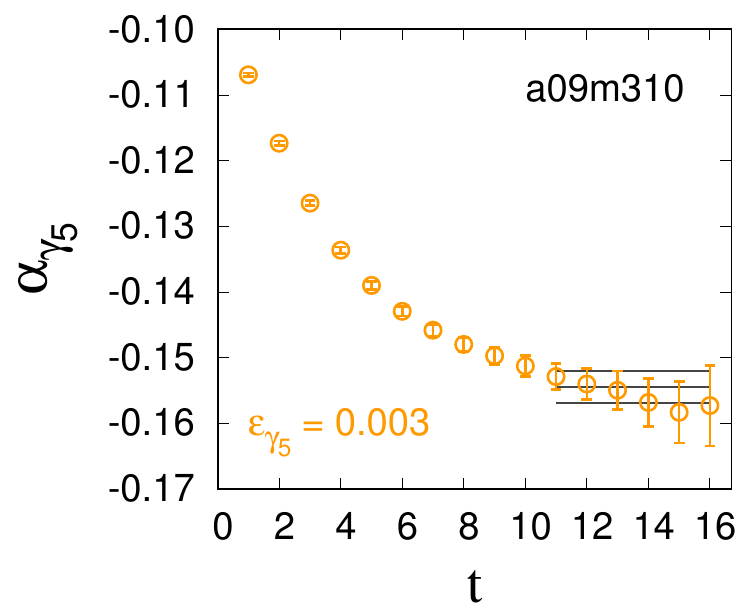} \quad
  \includegraphics[width=0.31\textwidth]{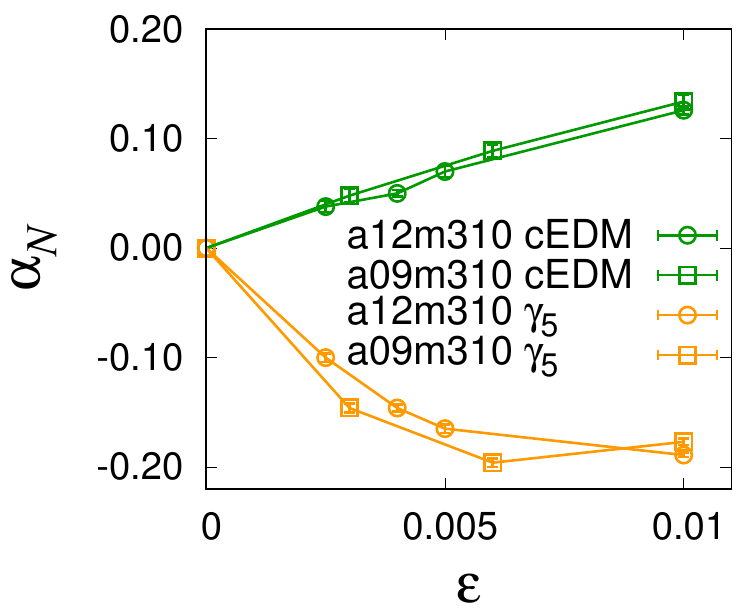} 
  \caption{The left panel shows the extraction of the CP violating
    phase $\alpha$ induced by the cEDM operator from the long-time
    behavior of the 2-point correlator.  The middle panel shows the
    extraction of the phase for the $\gamma_5$ operator insertion. The
    right panel shows that both kinds of phases are linear in
    $\varepsilon$ when $\varepsilon$ is small. These data are 
    from the Clover-on-HISQ analysis on the $M_\pi\approx 310$~MeV ensembles 
    with lattice spacing $a=0.12$ (a12m310) and $0.09$ (a09m310) fm.  }
  \label{fig:alpha}
\end{figure}

Current calculations are at the stage of demonstrating a signal in the
full evaluation of Figure~\ref{fig:cedm-swinger-diag} with both the
cEDM and the $\gamma_5$ operator with which it mixes under
renormalization as discussed in Section~\ref{sec:cEDMrenorm}. First,
in Figure~\ref{fig:alpha}, we show the success at extraction of the phase
$\alpha$ that is induced in the ground state nucleon spinor by the CP
violating cEDM and $\gamma_5$ operators.  In the right panel of
Figure~\ref{fig:alpha}, we also show that this $\alpha$ is linear in
$\varepsilon$, which allows us to tune the value of $\varepsilon$. We
want to use a large $\varepsilon$ to increase the signal but remain in
the linear response regime.

\begin{figure}[tb]
  \centering
  \includegraphics[width=0.43\textwidth]{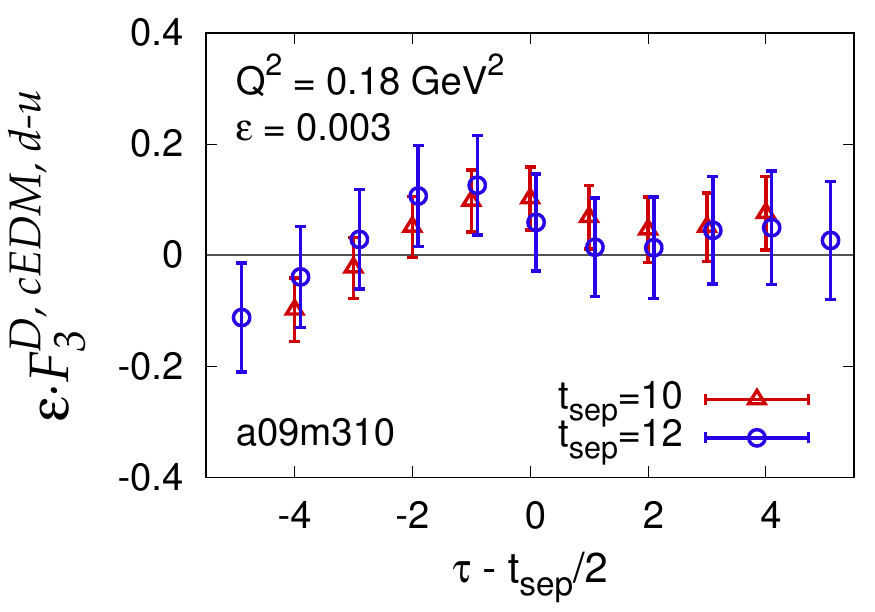} \qquad
  \includegraphics[width=0.43\textwidth]{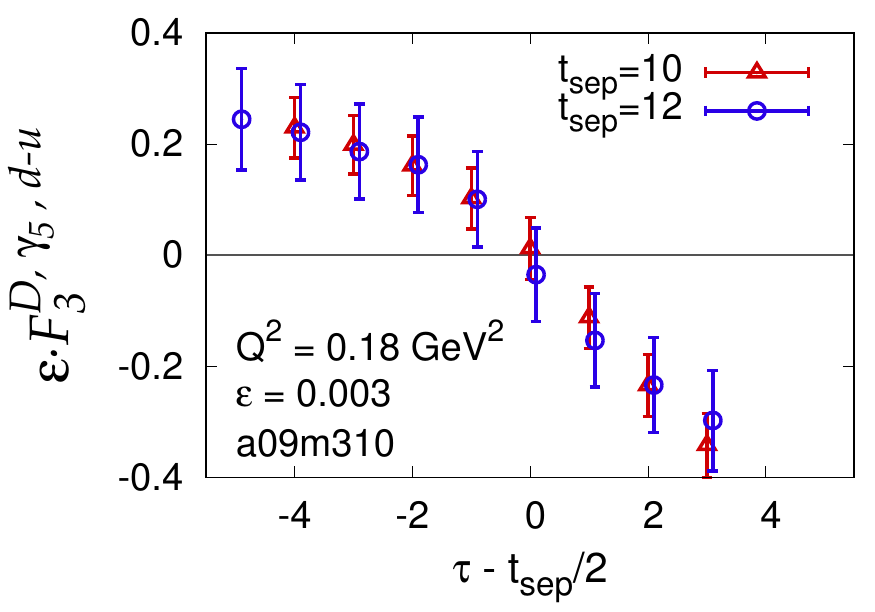}
  \caption{(Left) Signal of $F_3$ form-factor for two different separation of source
    and sink on a09m310 lattices. cEDM is inserted on d-quarks. Only
    the connected diagrams are considered. (Right) Same plot as the
    (middle), but $\gamma_5$ is inserted instead of the cEDM
    operator. Results shown are for the bare operators.}
  \label{fig:cedm-swinger-res}
\end{figure}

First examples of the quality of the signal in the contribution of the
connected diagrams to the form factor $F_3$ induced by the cEDM and
$\gamma_5$ operators are shown in Figures~\ref{fig:cedm-swinger-res}
for two different source and sink separations.  At present, the signal
is consistent with zero for both operators. 

\subsection{Renormalization of cEDM Operator}
\label{sec:cEDMrenorm}

The renormalization of the cEDM operator is studied both in one-loop
perturbation theory with Twisted-mass fermions
\cite{Constantinou:2015ela}, and in nonperturbative RI-SMOM scheme
\cite{Bhattacharya:2015rsa}. The most challenging outcome is the
divergent mixing with lower-dimensional operators: in particular the
$1/a^2$ mixing with the pseudoscalar quark bilinear operator:
\begin{align}
  O_\mathrm{cEDM} = a^2 \bar{q} \sigma^{\mu\nu}\gamma_5 G_{\mu\nu} q, \qquad O_P = \bar{q} \gamma_5 q\,.
\end{align}
Because the mixing is $1/a^2$ divergent, it needs to be calculated
precisely so that a finite operator can be constructed, ensemble by
ensemble, by subtracting the two terms. In
Ref.~\cite{Constantinou:2015ela}, the authors presented the results of the
$1/a^2$ mixing coefficients for the Twisted-mass fermions calculated
non-perturbatively and using the one-loop perturbation theory. In
Ref.~\cite{Bhattacharya:2015rsa}, authors defined a momentum
subtraction scheme, RI-$\tilde{\text{S}}$MOM, for the non-perturbative
renormalization of the cEDM operator on the lattice, and provided
one-loop matching coefficients to the $\overline{\text{MS}}$ scheme in
continuum limit.

\section{Summary}

In this talk, we reviewed the recent lattice QCD calculations of the
contribution of dimension four and five CP violating operators to the EDM of
the neutron. 

We reviewed the three approaches, external electric field method,
expansion in ${{\theta}}$, and simulation with imaginary ${{\theta}}$, 
used for the calculation of the QCD $\Theta$-term in Section~\ref{sec:theta}.
We also summarized the observation made in Ref.~\cite{Abramczyk:2017oxr}
concerning the subtlety of properly defining the neutron spinor in the
presence of parity violating interactions and its consequences
vis-a-vis mixing between the form factors $F_2$ and $F_3$ in
Section~\ref{sec:parity}.  All calculations based on extracting $F_3$,
previous to Ref.~\cite{Abramczyk:2017oxr}, had missed this
mixing. After correction, current lattice QCD results for the QCD
$\Theta$-term, are reduced by about a factor of ten and do not show a
statistically significant non-zero signal.

The quark EDM contribution to the neutron EDM can be written in terms
of the neutron tensor charges $g_q^T$. Currently, the $g_{u,d}^T$ are
determined to within 15\% uncertainty including the quark-line
disconnected diagrams, and $g_{s}^T$ is known to be very small. The
results are summarized in Section~\ref{sec:qedm}.

Lattice QCD calculation for the quark chromo-EDM operator have just
started.  Three methods for the calculation of the contribution of
cEDM and the construction of a finite renormalized operator are
described in Section~\ref{sec:cedm}.  The methodology for calculation
of the 3-point (or 4-point) correlation functions is under control but
a non-zero signal in all the diagrams has yet to be demonstrated. Once
a signal is achieved, we will still be a long way away from a full
calculation due to the divergent mixing between the cEDM and
$\gamma_5$ operators.

A few brief words on other lattice QCD calculations related to the
neutron EDM that have not been covered in this review. First, preliminary
results for the Weinberg three-gluon operator, one of the
dimension-six operators, were presented at the Lattice 2017 conference
\cite{Dragos:2017talk}. Calculations of the Weinberg operator are at the stage of
demonstrating a signal and have not even begun to address the issue of
renormalization that is potentially even more complicated than that
for the cEDM operator. Calculations exploring the four-fermion dimension 
six operators are yet to begin. Second, there are efforts to study the neutron 
EDM using lattice spectroscopy, and other matrix elements combined with 
chiral perturbation theory \cite{deVries:2016jox}, in addition to the direct
calculation of the matrix elements with the CP violating operators. 

To summarize, the theoretical calculations of matrix elements needed
to extend the importance and reach of nEDM experiments to constrain
BSM physics have begun in earnest but are very challenging and will
require many new innovations in both theory and computations.

\section*{Acknowledgments}

We thank Keh-Fei Liu, Gerrit Schierholz, Eigo Shintani, and Sergey
Syritsyn for valuable discussions and for sharing results. The work was
supported by the U.S. DoE HEP Office of Science contract number
DE-KA-1401020 and the LANL LDRD Program. The simulations for the qEDM
contribution and the cEDM contribution using Swinger source method are
carried out on computer facilities at (i) the USQCD Collaboration, 
which are funded by the Office of Science of the U.S. Department of Energy, 
(ii) the National Energy Research Scientific Computing Center, a DOE Office 
of Science User Facility supported by the Office of Science of the U.S. 
Department of Energy under Contract No. DE-AC02-05CH11231,
(iii) Oak Ridge Leadership Computing Facility at the Oak Ridge National 
Laboratory, which is supported by the Office of Science of the U.S. 
Department of Energy under Contract No. DE-AC05- 00OR22725, and
(iv) Institutional Computing at Los Alamos National Laboratory.

\bibliography{refs}

\end{document}